\documentclass[reprint,superscriptaddress, amsmath,amssymb,aps,notitlepage]{revtex4-1}
\usepackage{graphicx}
\usepackage{dcolumn}
\usepackage{bm}
\usepackage[colorlinks,linkcolor=blue,citecolor=blue]{hyperref}

\begin{document}

\title{Temperature Dependence of Polymer Network Diffusion}

\author{Takeshi Fujiyabu}
\thanks{T. Fujiyabu and T. Sakai contributed equally to this work.}
\affiliation{Department of Bioengineering, The University of Tokyo, 7-3-1 Hongo, Bunkyo-ku, Tokyo 113-8656, Japan.}
\author{Takamasa~Sakai}
\email[Corresponding author. ]
{sakai@tetrapod.t.u-tokyo.ac.jp}
\affiliation{Department of Bioengineering, The University of Tokyo, 7-3-1 Hongo, Bunkyo-ku, Tokyo 113-8656, Japan.}

\author{Ryota Kudo}
\affiliation{Department of Bioengineering, The University of Tokyo, 7-3-1 Hongo, Bunkyo-ku, Tokyo 113-8656, Japan.}
\author{Yuki Yoshikawa}
\affiliation{Department of Bioengineering, The University of Tokyo, 7-3-1 Hongo, Bunkyo-ku, Tokyo 113-8656, Japan.}
\author{Takuya Katashima}
\affiliation{Department of Bioengineering, The University of Tokyo, 7-3-1 Hongo, Bunkyo-ku, Tokyo 113-8656, Japan.}
\author{Ung-il~Chung}
\affiliation{Department of Bioengineering, The University of Tokyo, 7-3-1 Hongo, Bunkyo-ku, Tokyo 113-8656, Japan.}
\author{Naoyuki~Sakumichi}
\email[Corresponding author. ]
{sakumichi@tetrapod.t.u-tokyo.ac.jp}
\affiliation{Department of Bioengineering, The University of Tokyo, 7-3-1 Hongo, Bunkyo-ku, Tokyo 113-8656, Japan.}

\date{\today}

\begin{abstract}
The swelling dynamics of polymer gels are characterized by the (collective) diffusion coefficient $D$ of the polymer network.
Here, we measure the temperature dependence of $D$ of polymer gels with controlled homogeneous network structures using dynamic light scattering.
An evaluation of the diffusion coefficient at the gelation point $D_{\mathrm{gel}}$ and the increase therein as the gelation proceeds $\Delta D\equiv D-D_{\mathrm{gel}}$ indicates that $\Delta D$ is a linear function of the absolute temperature with a significantly large negative constant term.
This feature is formally identical to the recently discovered ``negative energy elasticity'' [Y.~Yoshikawa \textit{et al}., Phys. Rev. X \textbf{11}, 011045 (2021)], demonstrating a nontrivial similarity between the statics and dynamics of polymer networks.
\end{abstract}

\maketitle

\textit{Introduction}.---
The entropy elasticity of rubberlike solids (e.g., rubbers and polymer gels) and Brownian motion are described by formally analogous equations as both originate from thermal fluctuations.
In rubberlike solids, the shear modulus $G$ is conventionally considered to be proportional to the absolute temperature $T$, such that $G=nk_{B}T$ for an affine network model \cite{Flory1953}.
Here, $n$ is the number density of elastically effective strands, and $k_{B}$ is the Boltzmann constant.
Similarly, the Einstein relation \cite{Einstein1905} indicates that the (self) diffusion coefficient of Brownian motion $D_{s}$ is described as $D_{s}=\mu k_{B}T$, where $\mu$ is the mobility.
This Letter reports that the investigation of the above analogy led to the discovery of a nontrivial law that describes the temperature dependence of the (collective) diffusion coefficient $D$ of a polymer network in a solvent, which characterizes the swelling dynamics of a polymer gel.
Notably, the temperature dependence of $D$ is crucial for controlling the kinetics in applications involving sensors and actuators \cite{SensorActuator1,SensorActuator2}.

\begin{figure}[b!]
\centering
\includegraphics[width=\linewidth]{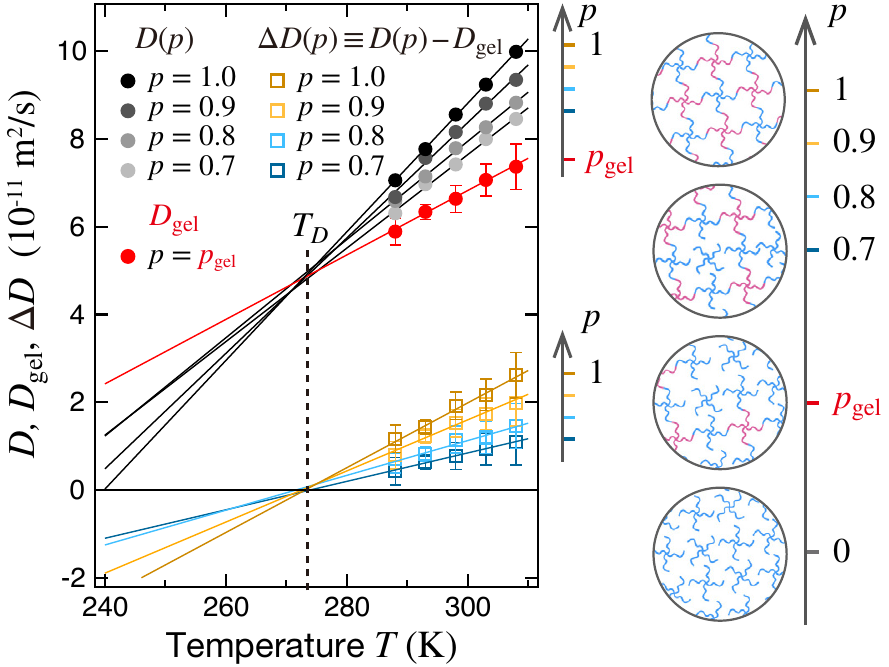}
\caption{
Temperature ($T$) dependence of (collective) diffusion coefficient $D$ and its components ($D_{\mathrm{gel}}$ and $\Delta D$) in four gel samples with different network connectivities ($p=0.7$, $0.8$, $0.9$, and $1.0$) for a polymer concentration of $c=60$~g/L.
(see Supplemental Material, Fig.~S1 for $c=30$, $90$, and $120$~g/L.)
The black and gray circles represent the experimental results of $D(T,p)$ for each sample. 
By extrapolating $D(T,p)$ to the gelation point ($p\to p_{\mathrm{gel}}$), we obtained $D_{\mathrm{gel}}(T)$ (red circles) [see Fig.~\ref{fig:2}(a)]. 
Then, we obtained $\Delta D$ as $\Delta D(T,p)=D(T,p)-D_{\mathrm{gel}}(T)$ (colored squares) for each sample.
The lines represent the least-squares fits of $D$, $D_{\mathrm{gel}}$, and $\Delta D$. 
We determine the temperature $T_{D}$ at which $\Delta D$ does not contribute to $D$, i.e., $D(T_{D}, p) = D_{\mathrm{gel}}(T_{D})$ and $\Delta D(T_{D}, p) = 0$. 
(Schematic illustration) 
The gel samples were synthesized by $AB$-type cross-end coupling of two precursors (tetra-arm polymers) to tune $p$ ($0\leq p\leq 1$) after the completion of the reaction, while maintaining $c$.
}
\label{fig:1}
\end{figure}

With regard to the elasticity of a polymer gel, we recently found that $G$ is not proportional to $T$ but is a linear function of $T$ with a significantly large negative constant term $b$ as $G=aT+b$ \cite{Yoshikawa2021}. 
Here, the negative value of $b$ is interpreted as ``negative energy elasticity'' because the first and second terms ($aT$ and $b$) correspond to the entropy and internal energy contributions to $G$, respectively. 
This result disproves the conventional assertion that the gel elasticity is approximately proportional to the absolute temperature (i.e., $G\simeq aT$), similar to the rubber elasticity.

Moreover, by examining more than 50 different polymer network structures, we found \cite{Yoshikawa2021} that $G$ is governed by
\begin{equation}
G(T,c,p)=a(c,p)\left[T-T_{0}(c)\right],
\label{eq:ElasticModulus}
\end{equation}
where $c$ is the polymer concentration and $p$ is the network connectivity characterizing the degree of gelation progress.
Here, we consider the as-prepared state of the entire system consisting of gel and sol components.
Because gelation is a dynamic process during which network connections are formed, we cannot precisely measure the physical properties at a certain point during gelation.
To statically replicate the gelation proceed at any point, we developed a methodology that enables measurements using $p$ as a control parameter \cite{Yoshikawa2019,Sakumichi2021}. 
This method utilizes $AB$-type cross-end coupling of two precursors (tetra-arm polymers) whose terminal functional groups ($A$ and $B$) are mutually reactive \cite{Sakai2008} (the schematics in Fig.~\ref{fig:1}).
By mixing the two precursors in stoichiometrically balanced and imbalanced ratios, we can tune $p$ after the completion of the reaction as $p\simeq 2s$, while maintaining $c$.
Here, $p$ ($0\leq p\leq 1$) is defined as the fraction of the reacted terminal functional groups to all the terminal functional groups, and $s$ ($0\leq s\leq 1/2$) is the molar mixing fraction of the precursors of the minor group.
Using this method, we experimentally obtained Eq.~(\ref{eq:ElasticModulus}), which implies that $T_{0}$, which governs negative energy elasticity, is independent of $p$.

This Letter presents that an equation formally identical to Eq.~(\ref{eq:ElasticModulus}) holds not only for elasticity but also for the elastic contribution of diffusion of the polymer network in a solvent.
The diffusion of a network is characterized by the (collective) diffusion coefficient $D$ \cite{THB1973,LiTanaka1990} and is represented by the diffusion equation for the displacement vector of the network $\mathbf{u}$ as
\begin{equation}
\frac{\partial}{\partial t} \mathbf{u}=D\nabla^2 \mathbf{u},
\label{eq:DiffusionEq}
\end{equation}
where $t$ is the time, and $\nabla^2$ is the Laplacian operator. [For gels with a general shape, there exists an additional term proportional to $\nabla\times(\nabla\times\mathbf{u})$ in Eq.~(\ref{eq:DiffusionEq}) \cite{LiTanaka1990}, but it is not relevant to this study.] 
Figure~\ref{fig:1} summarizes the main results.
Remarkably, at a fixed $c$, the extrapolations of the linear fits for the $T$ dependence of $D$ with varying values of $p$ intersect at a single point [that is, ($T_{D}, D(T_{D})$)].
To interpret the physical meaning of this intersection, we focus on the change in $D$ as the gelation proceeds, that is, the $p$ dependence of $D$.
Throughout this Letter, we only consider gel states, that is, $p_{\mathrm{gel}}<p<1$, where $p_{\mathrm{gel}}$ represents $p$ at the gelation point.
We decompose $D$ into two components: $D(T,c,p)=D_{\mathrm{gel}}(T,c)+\Delta D(T,c,p)$, where we define
\begin{equation}
D_{\mathrm{gel}}(T,c)\equiv \lim_{p\to p_{\mathrm{gel}}}D(T,c,p)
\label{eq:def:Dmix}
\end{equation}
and the increment from $D_{\mathrm{gel}}$ as gelation proceeds as
\begin{equation}
\Delta D(T,c,p)\equiv D(T,c,p)-D_{\mathrm{gel}}(T,c),
\label{eq:def:Del}
\end{equation}
which originates from the polymer network elasticity
(see the section ``\textit{Analysis based on the THB theory}'' below).
As shown in Fig.~\ref{fig:1}, we find that $\Delta D$ has a form equivalent to Eq.~(\ref{eq:ElasticModulus}) as
\begin{equation}
\Delta D (T,c,p)=a_D (c,p)\left[T-T_{D} (c)\right],
\label{eq:ElasticPart}
\end{equation}
where $T_{D}$ is independent of $p$.
This property is a nontrivial feature formally identical to $T_{0}$.
We experimentally confirmed Eq.~(\ref{eq:ElasticPart}) for various concentrations (Supplemental Material, Fig.~S1 for $c=30$, $60$, $90$, and $120$~g/L).

We emphasize that the procedure in Eqs.~(\ref{eq:def:Dmix})--(\ref{eq:ElasticPart}) is not based on any assumptions or models, but instead relies only on well-defined parameters to analyze $D$.
Many studies have investigated $D$ of polymer gels using light scattering measurements by tuning various parameters, such as 
the ionization degree \cite{Ilmain1989}, 
number of pendant chains \cite{Bastide1979}, 
polymer volume fraction \cite{Davidson1985}, 
cross-link density \cite{Shibayama1996a}, 
solvent quality \cite{Mccoy2010}, 
and temperature \cite{THB1973,Davidson1985,Mccoy2010}. 
Notably, most studies \cite{Bastide1979,Davidson1985,Shibayama1996a,Mccoy2010,Shibayama1996b} have assumed the Stokes-Einstein relation 
[$D_{s} = k_{B}T/(6\pi\eta\xi)$ with the solvent viscosity $\eta$ and the correlation length $\xi$]. 
However, the applicability of the Stokes-Einstein relation to polymer gels has not been strictly validated, and the meaning of $\xi$ is unclear.
In contrast, the Tanaka, Hocker, and Benedek (THB) theory \cite{THB1973} contains only well-defined parameters. 
In this Letter, we first analyze the experimental results in terms of the procedure in Eqs.~(\ref{eq:def:Dmix})--(\ref{eq:ElasticPart}) and then interpret our analysis based on the THB theory.\\

\begin{figure*}[t!]
\centering
\includegraphics[width=\linewidth]{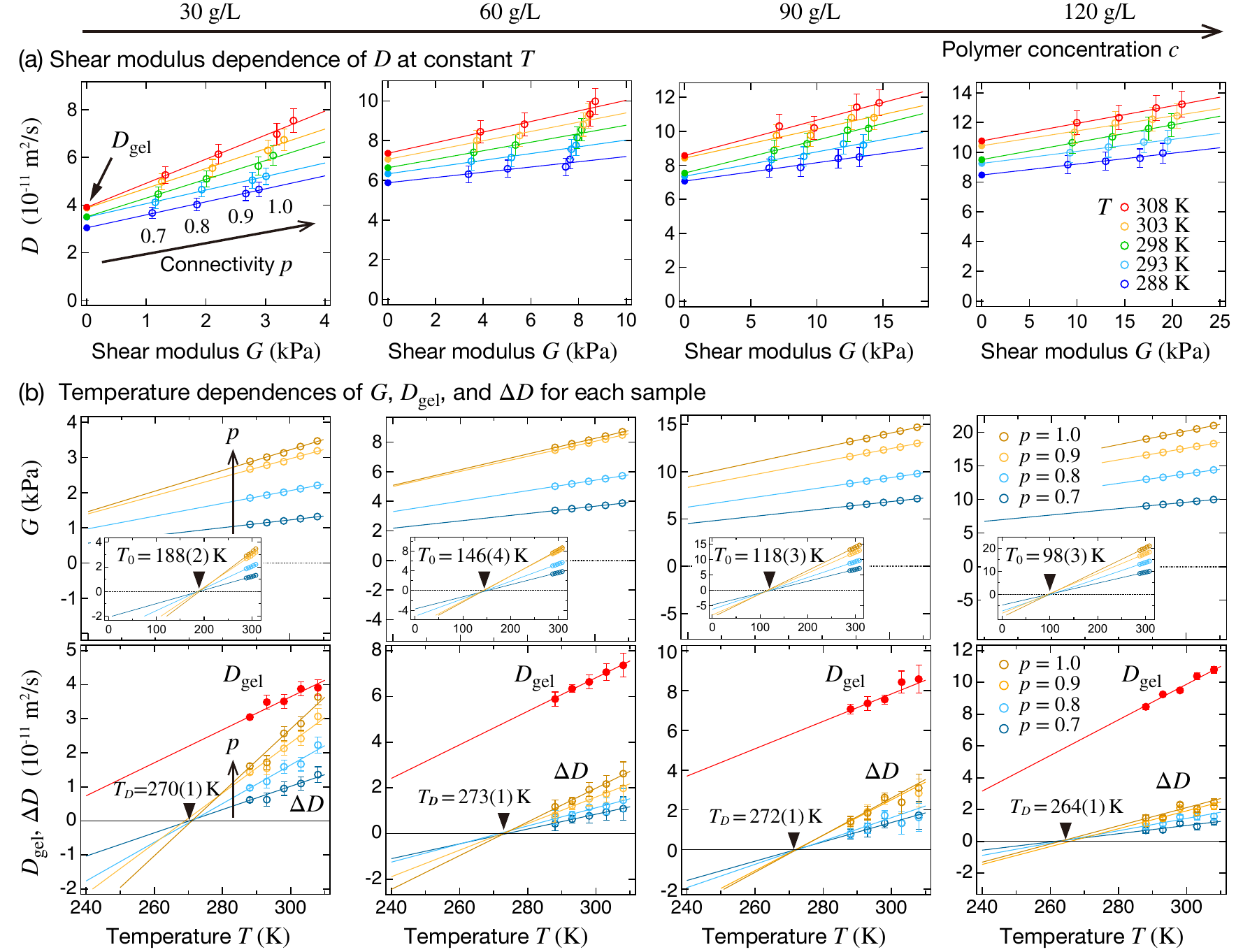}
\caption{
Formally identical relationships governing the shear modulus $G$ and $\Delta D$. 
We synthesized the gel samples with $16$ different network structures ($c=30$, $60$, $90$, $120$~g/L and $p=0.7$, $0.8$, $0.9$, $1.0$).
We measured $G$ and $D$ at $T=288$, $293$, $298$, $303$, and $308$~K.
The data of $G$ are partly taken from Ref.~\cite{Yoshikawa2021}.
(a) Shear modulus dependence of $D$ at constant $T$. 
The open circles represent the experimental results, and the lines represent the least-squares fits of $D=D(G)=D_{\mathrm{gel}} + \alpha G$.
Here, $D(G=0)$ corresponds to $D_{\mathrm{gel}}$ (filled circles) because of the definition in Eq.~(\ref{eq:def:Dmix}).
(b) Temperature dependence of $G$ (upper panels) and $D_{\mathrm{gel}}$ and $\Delta D\equiv D-D_{\mathrm{gel}}$ (lower panels).
Insets in the upper panels depict the same results in different ranges of $T$.
The open circles represent the experimental results, and the lines represent the least-squares fits of the $T$ dependences of $G$, $D_{\mathrm{gel}}$, and $\Delta D$. 
All extrapolations of $G = G(T)$ and $\Delta D = \Delta D(T)$ with the same $c$ pass through $T_{0}$ and $T_{D}$ on the $T$ axis, leading to Eqs.~(\ref{eq:ElasticModulus}) and~(\ref{eq:ElasticPart}), respectively.
The value of $T_{0}$ or $T_{D}$ in each graph represents the average of the four samples with different $p$, and the values in parentheses represent the standard deviation.
}
\label{fig:2}
\end{figure*}

\textit{Materials and methods}.---
For the model polymer gel with a homogeneous network structure, we used a tetra-arm poly(ethylene glycol) (tetra-PEG) hydrogel \cite{Sakai2008}, which is synthesized by $AB$-type cross-end coupling of two precursors: tetra-arm PEG units with molar masses of $M=20$~kg/mol (NOF Co., Japan and XIAMEN SINOPEG BIOTECH Co., Ltd., China).
Each end of the tetra-arm PEG is modified with mutually reactive maleimide (tetra-PEG-MA) and thiol (tetra-PEG-SH).
All other reagents were purchased from WAKO.
All the materials were used without further purification. 
To optimize the gelation time \cite{Kurakazu2010}, we dissolved each precursor in a phosphate-citric acid buffer (pH $3.8$), where the molar concentration was $68$ and $200$~mM for the dynamic light scattering (DLS) measurement and the dynamic viscoelasticity measurement, respectively.
We previously confirmed that the results of the latter measurement are almost independent of the molar concentration of the buffer \cite{Yoshikawa2021}.
The polymer concentrations ($c$) were $30$, $60$, $90$, and $120$~g/L. 
We mixed the precursor solutions in different proportions of $s=0.35$, $0.40$, $0.45$, and $0.50$, where $s=[\textrm{tetra-PEG-SH}] / ([\textrm{tetra-PEG-MA}] + [\textrm{tetra-PEG-SH}])$, to obtain gel samples with tuned $p$ after the completion of the reaction as $p \simeq 2s$ \cite{Yoshikawa2019}.
All gel samples were used in the as-prepared state.

We measured $D$ via DLS on an ALV/CGS-3 compact goniometer system (ALV, Langen, Germany) in the same way as in Refs.~\cite{Fujiyabu2019,Kim2020}.
The gel samples were fabricated in a DLS glass tube (disposable culture tube 9830-1007 with an inner diameter of $8.4$ mm; IWAKI, Japan).
We measured the scattered light intensity $I(t)$ at time $t$ at a scattering angle of $\pi/2$ at $288$, $293$, $298$, $303$, and $308$~K for $600$~s.
Then, we evaluated the autocorrelation functions $g^{(2)}(\tau) \equiv \left<I(0)I(\tau)\right>/\left<I(0)\right>^{2}$ for the delay time $\tau \simeq 0.01$--$0.1$~ms, corresponding to the concentration fluctuation of the polymer network \cite{Sakai2008,THB1973}.
Here, $\left<\cdots\right>$ denotes the time average.
Using $g^{(2)} (\tau)$ with the partial heterodyne model \cite{Pusey1989,Joosten1991,Shibayama2002}, we evaluate $D$, details of which appear in Sec.~S1 of Supplemental Material.

We measured the (static) shear modulus $G$ using a dynamic shear rheometer (MCR302, Anton Paar, Austria) in the same way as in Ref.~\cite{Yoshikawa2021}. 
The mixed solutions of tetra-PEG-MA and tetra-PEG-SH were poured into the interstice of the double cylinder of the rheometer. 
The time courses of the storage modulus $G'$ and loss modulus $G''$ were measured at 298 K with the applied shear strain $\gamma$ of 2.0\% and the angular frequency $\omega$ of $31$ rad/s.
After $G'$ reached the equilibrium, we measured the $\omega$ dependences of $G'$ and $G''$ at $288$, $293$, $298$, $303$, and $308$~K.
Here, we varied $\omega$ from 0.63 to 63 rad/s with $\gamma = 1.0\%$, which guarantees linear elasticity.
As $G'$ was independent of $\omega$ in this region (Supplemental Material, Fig.~S3), we considered $G'$ at $\omega=31$~rad/s and $\gamma = 1.0\%$ as $G$.

\newpage

\textit{Results and analysis}.---
We analyze the experimental results in terms of the procedure in Eqs.~(\ref{eq:def:Dmix})-(\ref{eq:ElasticPart}).
Figure~\ref{fig:2}(a) demonstrates that $D$ is a nearly linear function of $G$ for all values of $c$ and $T$.
Here, $G$ is controlled by tuning $p$ \cite{Yoshikawa2021,Yoshikawa2019,Yasuda2020,Sakumichi2021};
$G$ is an increasing continuous function of $p$, and $G\to 0$ for $p\to p_{\mathrm{gel}}$.
Based on the linearity and definitions of Eqs.~(\ref{eq:def:Dmix}) and~(\ref{eq:def:Del}), we determined $D_{\mathrm{gel}}(T,c)$ by extrapolating the $D$--$G$ relations to $G\to 0$ (corresponding to $p\to p_{\mathrm{gel}}$) and $\Delta D(T,c,p)\equiv D(T,c,p)-D_{\mathrm{gel}}(T,c)=\alpha(T,c)G(T,c,p)$, where $\alpha(T,c)$ is the slope.

\newpage

Figure~\ref{fig:2}(b) depicts the $T$ dependence of $G$, $D_{\mathrm{gel}}$, and $\Delta D$. 
Each of $G$ and $\Delta D$ is a linear function of $T$ with a significantly large negative constant term. 
The extrapolations of $\Delta D$ for each value of $c$ converge at $T = T_{D}$ on the $T$ axis. 
This unexpected law is described by Eq.~(\ref{eq:ElasticPart}), which is formally identical to Eq.~(\ref{eq:ElasticModulus}) for $G$ (meaning ``negative energy elasticity'' \cite{Yoshikawa2019}).
We also confirmed these results for similar gels with $M=10$ kg/mol (Supplemental Material, Fig.~S4).
We note that the actual value of $D$ would not follow the extrapolations at low $T$ away from the measured $T$ because the coexisting solvent (water) freezes at $T\simeq 273$~K.
Despite the similarity in the forms of Eqs.~(\ref{eq:ElasticModulus}) and~(\ref{eq:ElasticPart}), $T_{D}$ is significantly larger than $T_{0}$.
The implication of this difference is discussed in the next section.\\

\textit{Analysis based on the THB theory}.---
We interpret our experimental results based on the THB theory~\cite{THB1973}, which shows that $D=[K+(4/3)G]/f$ in Eq.~(\ref{eq:DiffusionEq}) for $p> p_{\mathrm{gel}}$. 
Here, $K \equiv c\partial \Pi/\partial c$ is the osmotic bulk modulus, $\Pi =\Pi_{\mathrm{mix}} + \Pi_{\mathrm{el}}$ is the total swelling pressure ($\Pi_{\mathrm{mix}}$ and $\Pi_{\mathrm{el}}$ are the mixing and elastic contributions, respectively), and $f$ is the friction coefficient (per unit volume) between the polymer and the solvent.
We cannot directly compare our experimental results with $D = [K+(4/3)G]/f$ because $K = K_{\mathrm{mix}} + K_{\mathrm{el}}$ consists of both the polymer-solvent mixing contribution ($K_{\mathrm{mix}}\equiv c\partial \Pi_{\mathrm{mix}}/\partial c$)
 and elastic contribution 
 ($K_{\mathrm{el}}\equiv c\partial \Pi_{\mathrm{el}}/\partial c$).
Assuming that (i) $\Pi_{\mathrm{el}}=-G$ \cite{Yasuda2020}, (ii) the scaling law $G \sim c^{\beta}$ during the swelling \cite{Obukhov1994}, and (iii) $f(T,c,p)\simeq f(T,c)$ (the water permeation experiments \cite{TokitaTanaka1991,Fujiyabu2017} confirm that the $p$ dependence of $f$ is quite weak), we can derive \cite{Fujiyabu2019,Kim2020}
\begin{equation}
D(T,c,p)\simeq\frac{K_{\mathrm{mix}}(T,c)}{f(T,c)}+\left(4/3-\beta\right)\frac{G(T,c,p)}{f(T,c)},
\label{eq:THB}
\end{equation}
where $\beta$ ranges from $1/3$ to $(9\nu -4)/(9\nu -3) \simeq 0.563$ depending on the polymer concentration of the gel. 
Here, $\nu \simeq 0.5876$ is the universal critical exponent for polymer solutions (or the self-avoiding walk) \cite{clisby2010accurate,clisby2016high,Kompaniets2017}.
In the gel state ($p > p_{\mathrm{gel}}$), $K_{\mathrm{mix}}$ and $\Pi_{\mathrm{mix}}$ are independent of $p$ (see Fig.~2 in Ref.~\cite{Yasuda2020}).
Equation~(\ref{eq:THB}) indicates that $D$ is a linear function of $G$ with increasing $p$, which is consistent with Fig.~\ref{fig:2}(a). 
Substituting Eq.~(\ref{eq:THB}) into Eq.~(\ref{eq:def:Dmix}), we have $D_{\mathrm{gel}}\simeq K_{\mathrm{mix}}/f$ because $G(T,c,p_{\mathrm{gel}})=0$.
Also, combining Eqs.~(\ref{eq:def:Del}) and~(\ref{eq:THB}), we obtain $\Delta D\simeq (4/3-\beta )G/f$.
Thus, $D_{\mathrm{gel}}$ and $\Delta D$ largely correspond to the mixing and elastic contributions to $D$, respectively.

\begin{figure}[t!]
\centering
\includegraphics[width=\linewidth]{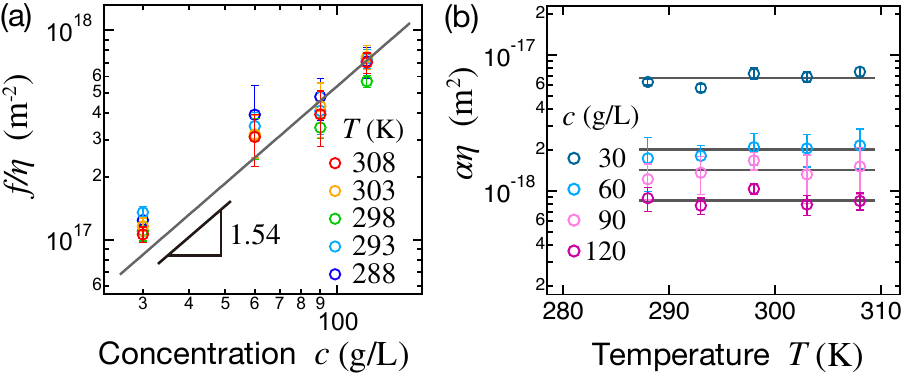}
\caption{(a) Log-log plots of the $c$ dependence of $f/\eta$, indicating $f/\eta \sim c^{2\nu/(3\nu -1)} \simeq c^{1.54}$ for $\nu\simeq 0.5876$.
Here, each friction coefficient between the polymer and the solvent $f$ is obtained from the linear fits in Fig.~\ref{fig:2}(a) by assuming Eq.~(\ref{eq:THB}) and $\beta\simeq 0.563$.
(b) Log plots of the $T$ dependence of $\alpha\eta$.
The viscosity of the solvent (water) $\eta$ is taken from Ref.~\cite{Kestin1978}.
The solid lines serve as a guide to the eye.
}
\label{fig:3}
\end{figure}

Using Eq.~(\ref{eq:THB}) with assuming $\beta\simeq 0.563$ (i.e., $4/3-\beta\simeq 0.770$) based on Ref.~\cite{Kim2020}, we evaluate $f$ and $K_{\mathrm{mix}}$ from the slopes [$(4/3-\beta ) /f$] and ordinate intercepts ($K_{\mathrm{mix}}/f$) of the linear fits in Fig.~\ref{fig:2}(a) \cite{Fujiyabu2019,Kim2020}.
Figure~\ref{fig:3}(a) shows that the obtained values of $f$ are consistent with the scaling relationships for semidilute solutions in a good solvent \cite{TokitaTanaka1991} $f/\eta \sim c^{2\nu/(3\nu -1)}$, where $\eta$ is the solvent viscosity.
Similarly, those of $K_{\mathrm{mix}}$ are consistent with $K_{\mathrm{mix}}/k_{B}T \sim c^{3\nu/(3\nu -1)}$ \cite{Cloizeaux1975,deGennes1979} (Supplemental Material, Fig.~S5).
Moreover, Fig.~\ref{fig:3}(a) indicates that $f/\eta$ is almost independent of $T$, which was experimentally shown using poly(acrylamide) gels (Fig.~9 in Ref.~\cite{TokitaTanaka1991}).
Therefore, our observations and analyses are consistent with the THB theory and scaling relationships.

To elucidate the origin of Eq.~(\ref{eq:ElasticPart}), we focus on the temperature ($T$) dependence of $\Delta D$.
Inspired by the THB theory, we plot the experimentally obtained $\alpha\eta$ against $T$ in Fig.~\ref{fig:3}(b) to show that the $T$ dependence of $\alpha\eta$ is quite weak.
Thus, we can set $\alpha\eta \simeq h(c)$ and obtain $\alpha (T,c)\simeq h(c)/\eta (T)$.
From the definition of $\alpha$, we have
\begin{equation}
\Delta D (T,c,p)=\alpha (T,c)G(T,c,p) 
\simeq
h(c) \frac{G(T,c,p)}{\eta(T)},
\label{eq:TemperatureDependenceDeltaD}
\end{equation}
showing that the $T$ dependence of $\Delta D$ is determined by $G/\eta$.
Because the $T$ dependence of $1/\eta$ is greater than that of $G$ (see Supplemental Material, Fig.~S6), $1/\eta$ almost determines the $T$ dependence of $\Delta D$.
Moreover, substituting Eq.~(\ref{eq:ElasticModulus}) into Eq.~(\ref{eq:TemperatureDependenceDeltaD}), we obtain Eq.~(\ref{eq:ElasticPart}) by neglecting the second order terms of $T$ (see Sec.~S2 of Supplemental Material).
Therefore, $T_{D}$ is significantly larger than $T_{0}$, and exhibits almost no concentration dependence.\\

\textit{Concluding remarks}.---
We experimentally investigated the temperature ($T$) dependence of the (collective) diffusion coefficient $D$ of polymer gels.
In Eqs.~(\ref{eq:def:Dmix}) and (\ref{eq:def:Del}), we operationally defined $D_{\mathrm{gel}}$ and $\Delta D$, which largely correspond to the mixing and elastic contributions, respectively [Eq.~(\ref{eq:THB})].
As depicted in Figs.~\ref{fig:1} and \ref{fig:2}(b), $\Delta D$ is a linear function of $T$ with a significantly large negative constant term, which is formally identical to that of the shear modulus \cite{Yoshikawa2021}.
At a certain temperature $T_{D}$, $\Delta D$ vanishes, and $D$ is independent of the network connectivity $p$ [Eq.~(\ref{eq:ElasticPart}) and Figs.~\ref{fig:1} and \ref{fig:2}(b)].
This simple unexpected law [Eq.~(\ref{eq:ElasticPart})] has not been predicted by any existing theory and can stimulate experimental and theoretical research of $T_{D}$.

Our findings demonstrate a nontrivial similarity between the statics and dynamics of polymer networks and provide new insights into the so-called diffusio-mechanical (or stress-diffusion) coupling \cite{FujineTakigawaUrayama2015,YamamotoMasubuchiDoi2018,Doi2021,ManDoi2021}.
Furthermore, our findings are important for controlling the swelling response time of stimuli-responsive gels, such as sensors and actuators \cite{SensorActuator1,SensorActuator2}.
This is because a temperature change of $20$~K can nearly double $D$ up to a maximum [$c=30$ g/L with $p=1$ in Fig.~\ref{fig:2}(a)], and this change is mainly caused by the elastic contribution $\Delta D$, which has not been explicitly discussed \cite{Bastide1979,Davidson1985,Shibayama1996a,Shibayama1996b,Shibayama2002} until recently \cite{Fujiyabu2019,Kim2020}.

\begin{acknowledgments}
This work was supported by the Japan Society for the Promotion of Science (JSPS) through Grants-in-Aid for
 JSPS Fellow Grant No.~19J22561 to T.F. and No.~21J13478 to Y.Y., 
 Scientific Research (A) Grant No.~21H04688 to T.S. and No.~21H04952 to U.C., 
 Transformative Research Areas Grant No.~20H05733 to T.S., 
and Early Career Scientists Grant No.~20K15338 to T.K. and No.~19K14672 to N.S.
This work was also supported by the Japan Science and Technology Agency (JST) through CREST Grant No.~JPMJCR1992 to T.S. and COI Grant No.~JPMJCE1304 to U.C.
\end{acknowledgments}


\end{document}


\title{Supplemental Material for:
``Temperature Dependence of Polymer Network Diffusion''}

\author{Takeshi Fujiyabu}
\thanks{T. Fujiyabu and T. Sakai contributed equally to this work.}
\affiliation{Department of Bioengineering, The University of Tokyo, 7-3-1 Hongo, Bunkyo-ku, Tokyo 113-8656, Japan.}
\author{Takamasa~Sakai}
\email[Corresponding author. ]
{sakai@tetrapod.t.u-tokyo.ac.jp}
\affiliation{Department of Bioengineering, The University of Tokyo, 7-3-1 Hongo, Bunkyo-ku, Tokyo 113-8656, Japan.}

\author{Ryota Kudo}
\affiliation{Department of Bioengineering, The University of Tokyo, 7-3-1 Hongo, Bunkyo-ku, Tokyo 113-8656, Japan.}
\author{Yuki Yoshikawa}
\affiliation{Department of Bioengineering, The University of Tokyo, 7-3-1 Hongo, Bunkyo-ku, Tokyo 113-8656, Japan.}
\author{Takuya Katashima}
\affiliation{Department of Bioengineering, The University of Tokyo, 7-3-1 Hongo, Bunkyo-ku, Tokyo 113-8656, Japan.}
\author{Ung-il~Chung}
\affiliation{Department of Bioengineering, The University of Tokyo, 7-3-1 Hongo, Bunkyo-ku, Tokyo 113-8656, Japan.}
\author{Naoyuki~Sakumichi}
\email[Corresponding author. ]
{sakumichi@tetrapod.t.u-tokyo.ac.jp}
\affiliation{Department of Bioengineering, The University of Tokyo, 7-3-1 Hongo, Bunkyo-ku, Tokyo 113-8656, Japan.}

\date{\today}

\maketitle

\begin{figure*}[h!]
\centering
\includegraphics[width=\linewidth]{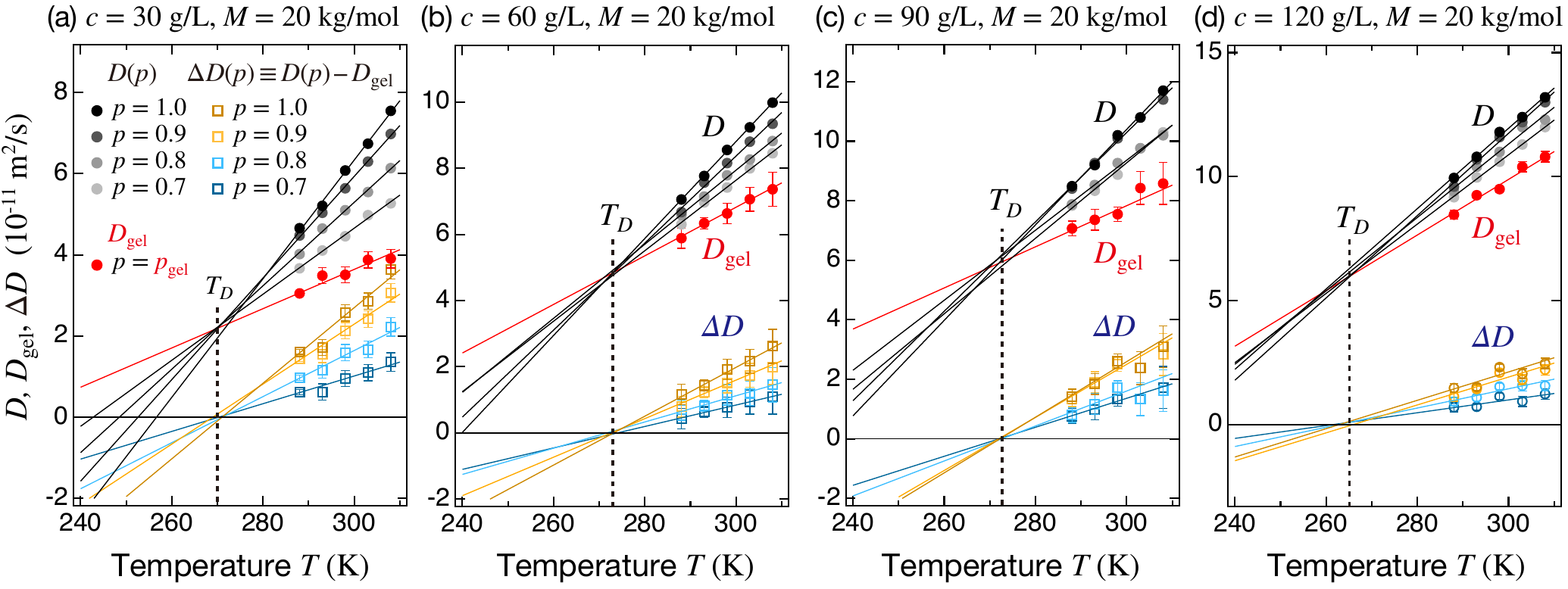}
\caption{
Temperature ($T$) dependence of collective diffusion coefficient $D$ and its components ($D_{\mathrm{gel}}$ and $\Delta D$) in $16$ samples of tetra-PEG hydrogels with different connectivities ($p = 0.7$, $0.8$, $0.9$, and $1.0$) for polymer concentrations of $c = 30$ (a), $60$ (b), $90$ (c), and $120$~g/L (d). 
Here, panel (b) is identical to the left panel in Fig.~1 in the main text. 
The black and gray circles represent the experimental results of $D(T,p)$ for each sample.
We obtained $D_\mathrm{mix}(T)$ (red circles) by extrapolating $D(T,p)$ at the gelation point (see Fig. 2a in the main text) and then obtained $D_\mathrm{el}$ as $D_\mathrm{el}(T,p) = D(T, p)- D_\mathrm{mix}(T)$ (colored squares).
The lines represent the least-squares fits of $D$, $D_\mathrm{mix}$, and $D_\mathrm{el}$ for each sample.
We determine the temperature $T_{D}$ at which $D_\mathrm{el}$ does not contribute to $D$, i.e., $D(T_{D}, p) = D_\mathrm{mix}(T_{D})$ and $D_\mathrm{el}(T_{D})=0$.
}
\label{sup:fig:S1}
\end{figure*}

\section{Evaluation of collective diffusion coefficient from autocorrelation function}

This section explains the approach to evaluate $D$ from the autocorrelation function
$g^{(2)}(\tau) \equiv \left<I(0)I(\tau)\right>/\left<I(0)\right>^{2}$
for the scattered light intensity $I(t)$ obtained by measuring the dynamic light scattering (DLS).
Here, $\left<\cdots\right>$ denotes the time average and we varied the delay time $\tau \simeq 0.01$--$0.1$~ms, corresponding to the concentration fluctuation of the polymer network.
A He-Ne laser with a power of $22$ mW emitting polarized light was used.
The scattering vector is $q = (4\pi n/\lambda)\sin(\theta/2) = 0.0187$~nm$^{-1}$, where $n= 1.333$ is the refractive index of the solvent, $\lambda = 632.8$~nm is the wavelength of the scattered light, and $\theta= \pi/2$ is the scattering angle.\\

\begin{figure*}[h!]
\centering
\includegraphics[width=0.47\linewidth]{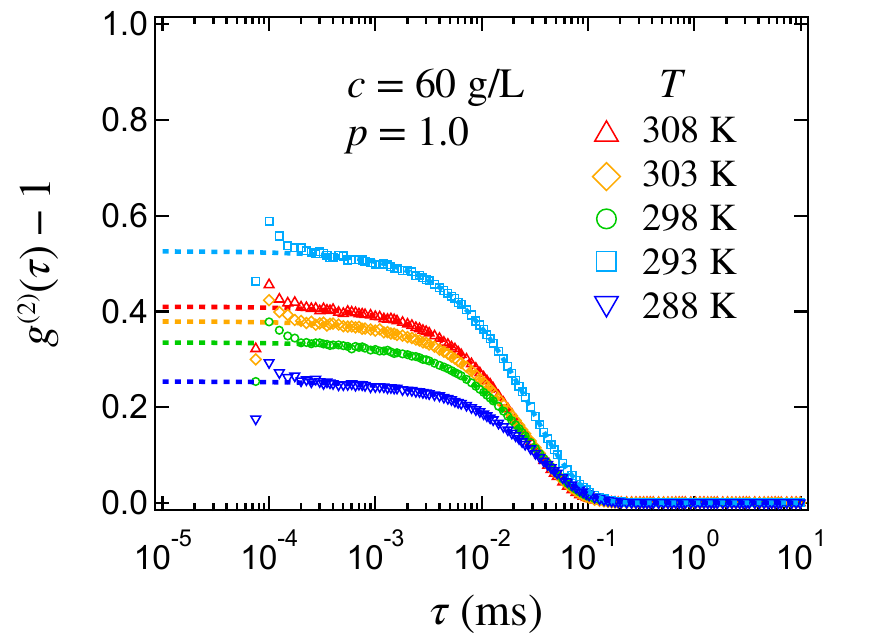}
\caption{
Autocorrelation functions ($g^{(2)}(\tau)-1$) of a gel sample with $c=60$~g/L, $p=1.0$, and $M=20$ kg/mol at various temperatures [$T = 308$ (triangles), $303$ (rhombuses), $298$ (circles), $293$ (squares), and $288$~K (inverted triangles)].
The dashed lines are the least-squares fits of the stretched exponential function given in Eq.~(\ref{sup:eq:g2}).
}
\label{sup:fig:g2}
\end{figure*}

We fitted the autocorrelation functions $g^{(2)}(\tau)$ using a stretched exponential function as
\begin{equation}
g^{(2)}(\tau)-1=A \exp\left[-2\left(\frac{\tau}{\tau_g} \right)^{\beta_{\mathrm{st}}} \right]+\epsilon,
\label{sup:eq:g2}
\end{equation}
where $A$ is the initial amplitude, $\tau_{g}$ is the relaxation time, 
$\beta_{\mathrm{st}}$ is the stretched exponent corresponding to the distribution of the network dynamics, and $\epsilon$ is the time-independent background. 
We show the typical fitting results in Fig.~\ref{sup:fig:g2} and Table~\ref{table}.
Figure~\ref{sup:fig:g2} indicates that Eq.~(\ref{sup:eq:g2}) is consistent with $g^{(2)}(\tau)$ of tetra-PEG hydrogels.
As for the fitting parameters ($A$, $\tau_{g}$, $\beta_{\mathrm{st}}$, $\epsilon$), provided in Table~\ref{table}, we observed two characteristic tendencies.
First, $\beta_{\mathrm{st}}\simeq 1.0$ suggests that the network dynamics are nearly homogeneous~\cite{Li2017}. 
Second, $0\leq\epsilon\ll A$ suggests that the effect of $\epsilon$ on the estimation of $\tau_{g}$ is negligible. 
These two tendencies indicate that the estimation of $D$ is independent of $\beta_{\mathrm{st}}$ and $\epsilon$ in our experiment.
The diffusion coefficient $D$ of ergodic mediums ($A\simeq 1.0$) is independent of the sample position and is estimated as $D = 1/q^{2}\tau_{g}$. 
In contrast, $A$ and $\tau_{g}$ of non-ergodic polymer gels ($A < 1.0$) depend on the sample position, resulting in a position-dependent diffusion coefficient $D_{A} = 1/q^{2}\tau_{g}$ (e.g., Ref.~\cite{Shibayama1996a}).\\

The partial heterodyne model \cite{Pusey1989,Joosten1991,Shibayama2002} enables us to estimate the position-independent $D$ of non-ergodic polymer gels. The light scattering from polymer gels includes two contributions from the thermal fluctuations and from the static (macroscopic) inhomogeneities in the polymer concentration. In the partial heterodyne model, $D$ is estimated as the thermal fluctuating component of the polymer concentration using the following equation
\begin{equation}
D=(2-X) D_A,
\label{sup:eq:DDA}
\end{equation}
where $X\equiv I_F/\left<I(0)\right>$ is the ratio of the intensity from the thermal fluctuation $I_F$ to the total intensity $\left<I(0)\right>$.
Assuming that the coherence factor is equal to $A$, we obtain $X=1-\sqrt{1-A}$.
Therefore, using Eq.~(\ref{sup:eq:DDA}) and $D_{A} = 1/q^{2}\tau_{g}$, we obtain
\begin{equation}
D=\frac{1+\sqrt{1-A}}{q^2 \tau_{g}}.
\label{sup:eq:D}
\end{equation}
The value of $D$ estimated using Eq.~(\ref{sup:eq:D}) is position independent and depends only on the parameters that characterize the condition of the gel network ($T, c, p$).\\

\begin{table}[h!]
\caption{Fitting parameters of least-squares fits of Eq.~(\ref{sup:eq:g2}) in Fig.~\ref{sup:fig:g2}.}
\label{table}
\begin{tabular}{ccccc}
\hline
\hline
$T$ (K) & $A$ & $\tau_{g}$ (ms) & \, $\beta_{\mathrm{st}}$ &  $\epsilon\quad$  \\
\hline
  $\qquad308\qquad$ & $\quad0.41\quad$ & $\quad0.051\quad$ & $\quad0.95\quad$ & $\quad0.00066\quad$ \\
  $303$ & $0.38$ & $0.055$ & $0.94$ & $0.00030$ \\
  $298$ & $0.33$ & $0.061$ & $0.95$ & $0.00089$ \\
  $293$ & $0.53$ & $0.063$ & $0.93$ & $0.00056$ \\
  $288$ & $0.25$ & $0.075$ & $0.95$ & $0.00029$ \\
  \hline
  \hline
\end{tabular} 
 \end{table}

\begin{figure*}[t!]
\centering
\includegraphics[width=0.62\linewidth]{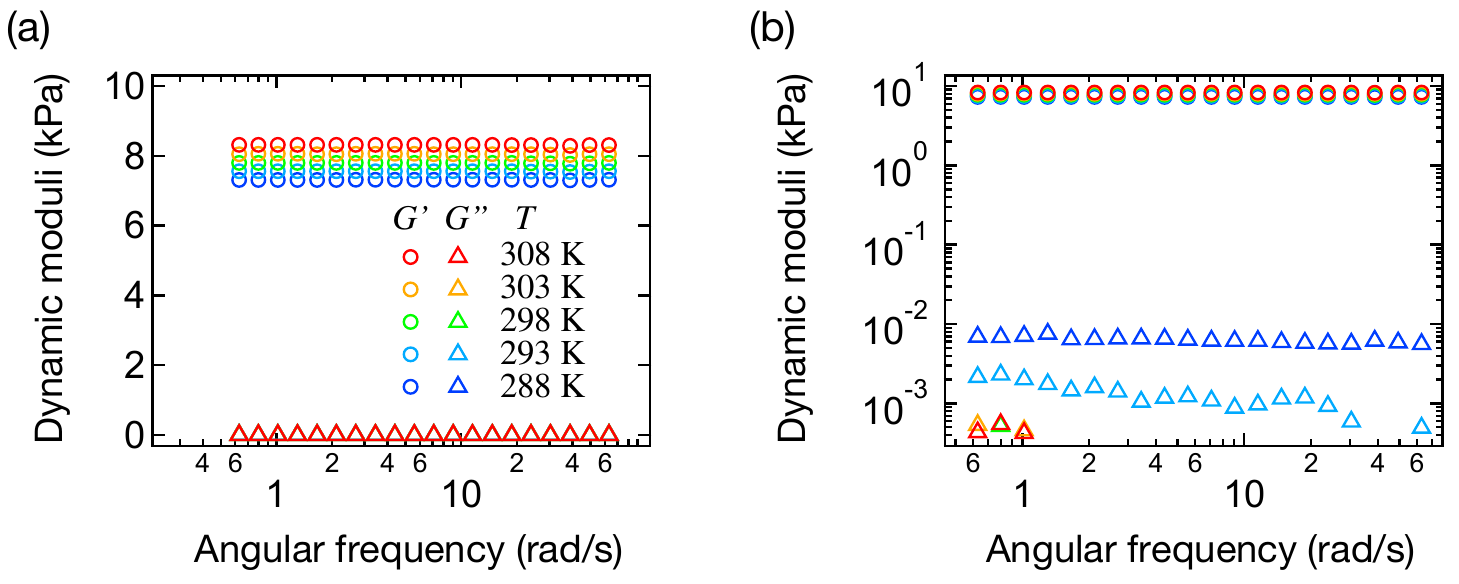}
\caption{
Linear-log (a) and log-log (b) plots of typical results of the angular frequency dependence of $G'$ (circles) and $G''$ (triangles) of a gel sample at different temperatures ($T = 308$, $303$, $298$, $293$, and $288$~K). 
The polymer concentration is $c=60$~g$/$L, connectivity is $p=1.0$, and the molar mass of the precursors is $M=20$~kg/mol.
}
\label{fig:concentration}
\end{figure*}

\begin{figure*}[h]
\centering
\includegraphics[width=0.67\linewidth]{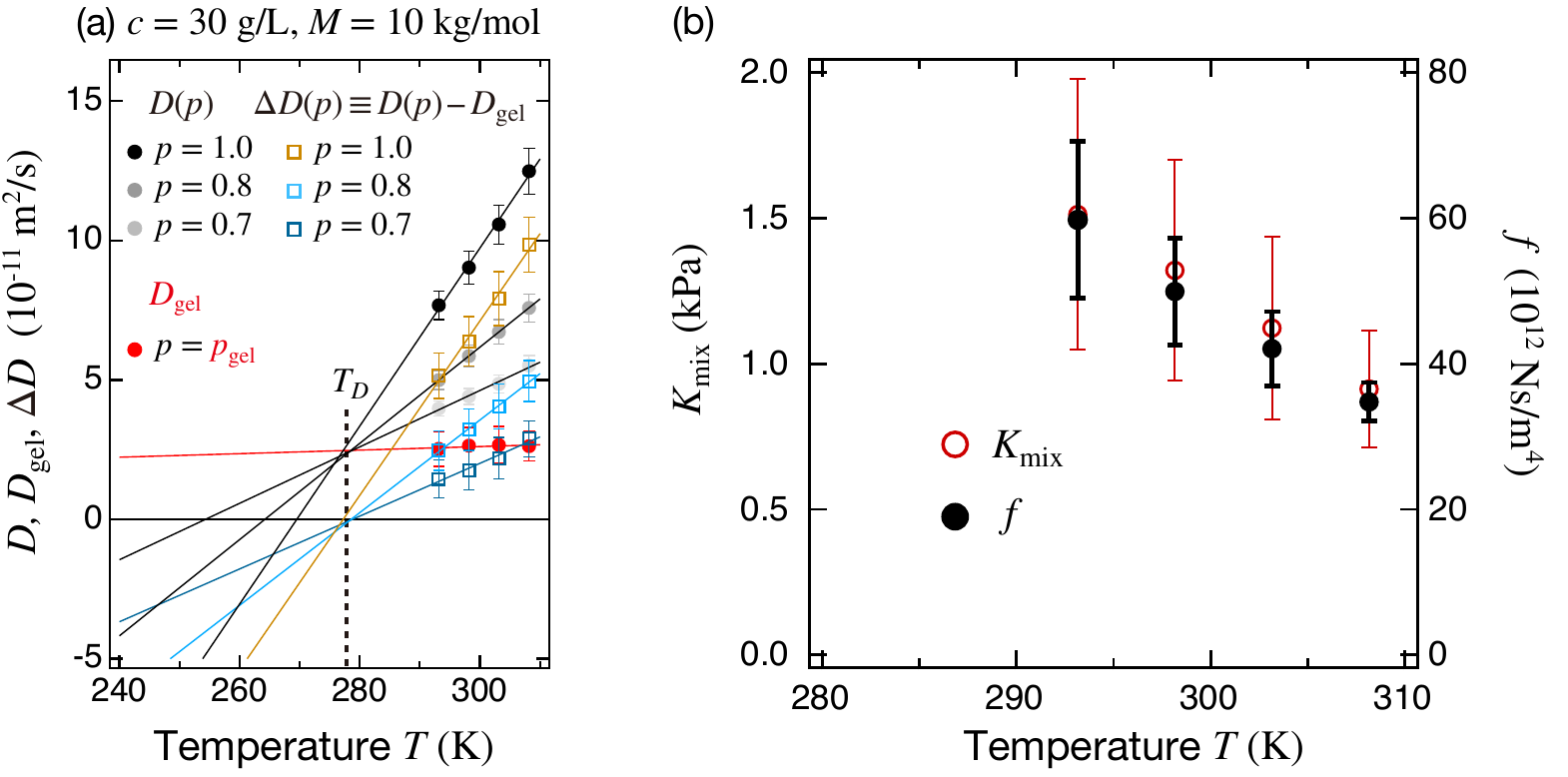}
\caption{
(a) Temperature ($T$) dependence of collective diffusion coefficient $D$ and its components ($D_{\mathrm{gel}}$ and $\Delta D$) in three samples of tetra-PEG hydrogels with different connectivities ($p = 0.7$, $0.8$, and $1.0$) for a polymer concentration of $c = 30$~g/L and molar mass of $M=10$~kg/mol.
The notations are the same as Fig.~\ref{sup:fig:S1}.
(b) Temperature ($T$) dependences of the polymer-solvent mixing contribution to the osmotic bulk modulus ($K_{\mathrm{mix}}$, open circles) and the friction coefficient (per unit volume) between the polymer and the solvent ($f$, filled circles) in tetra-PEG hydrogels with a polymer concentration of $c = 30$~g/L and molar mass of $M=10$~kg/mol.
Because of the coincidence of the temperature dependence trends of $K_{\mathrm{mix}}$ and $f$, $D_{\mathrm{gel}}$ in panel (a) is almost temperature independent.
}
\end{figure*}

\begin{figure*}[h!]
\centering
\includegraphics[width=0.79\linewidth]{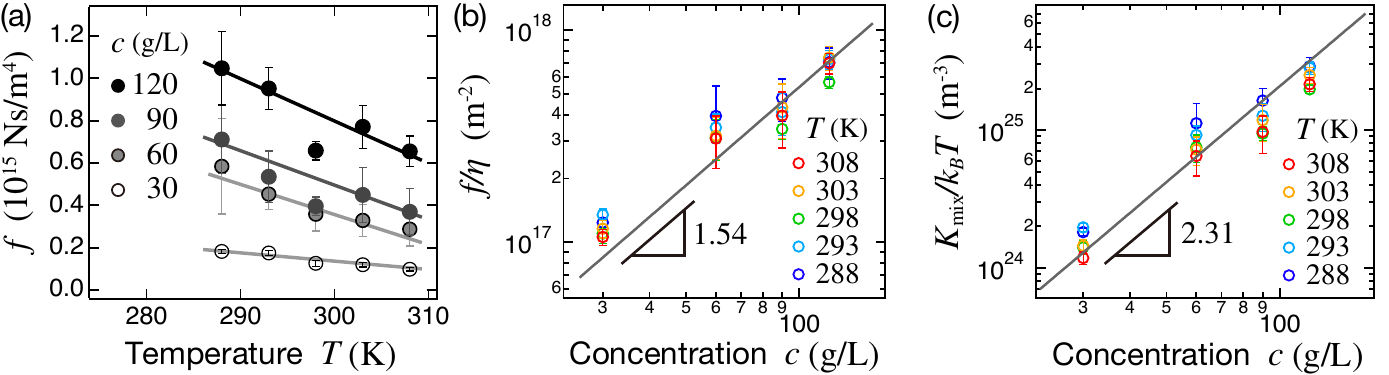}
\caption{
Friction coefficient between the polymer and the solvent f and the mixing contribution to the osmotic bulk modulus $K_{\mathrm{mix}}$ obtained from the linear fits in Fig. 2(a) in the main text. 
Here, we assume Eq. (6) in the main text with $\beta=0.563$ (i.e., $4/3-\beta =0.770$).
(a) The temperature dependence of f for different concentrations c. The solid lines serve as a guide to the eye. For all values of c, we find that f is a decreasing function of $T$, which is consistent with Ref.~\cite{TokitaTanaka1991}. 
(b-c) 
Log-log plot of the $c$ dependence of (b) $f/\eta$ and (c)  $K{\mathrm{mix}}/k_{B}T$. 
These results are consistent with the scaling relationships for semidilute solutions in a good solvent: $K_{\mathrm{mix}}/k_{B}T \sim c^{3\nu/(3\nu-1)} \simeq c^{2.31}$ and 
$f/\eta \sim c^{2\nu/(3\nu-1)} \simeq c^{1.54}$ for $\nu \simeq 0.5876$.
The value of the viscosity of the solvent (water) $\eta$ is taken from Ref.~\cite{Kestin1978}.
Panel (b) is identical to Fig.~3(a) in the main text.
}
\label{fig:concentration}
\end{figure*}

\section{On the origin of Equation~(5)}

\begin{figure*}[h]
\centering
\includegraphics[width=0.44\linewidth]{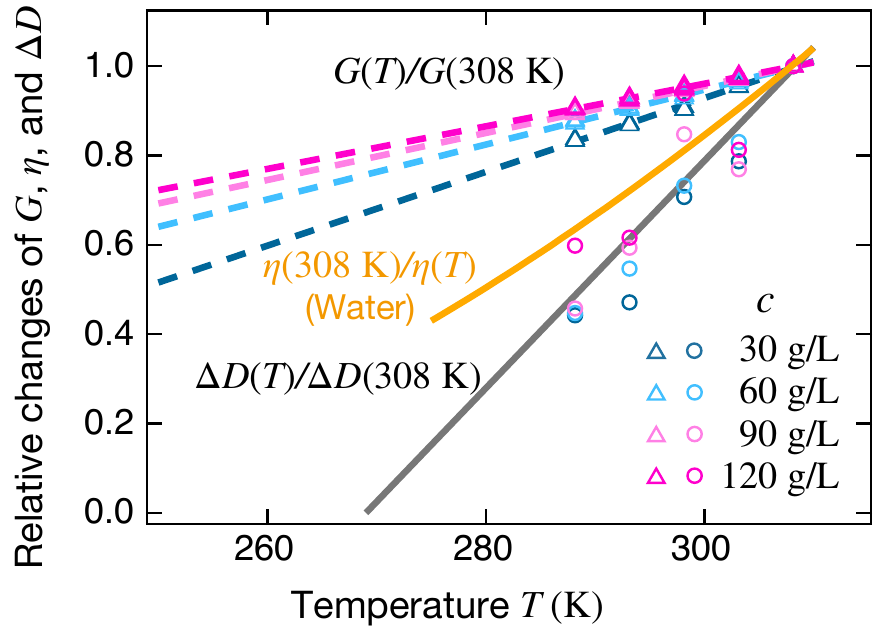}
\caption{
Temperature ($T$) dependence of the inverse of the viscosity of water 
$\eta(308\,\mathrm{K})/\eta(T)$ (orange curve), 
$G(T)/G(308\,\mathrm{K})$ (triangles), and 
$\Delta D(T)/\Delta D(308\,\mathrm{K})$ (circles) for polymer concentrations of $c=30$, $60$, $90$, and $120$~g$/$L.
The dashed lines represent the least-squares fits of the $T$ dependence of $G(T)/G(308\,\mathrm{K})$.
The gray solid line serves as a guide to the eye for $\Delta D(T)/\Delta D(308\,\mathrm{K})$.
The value of the viscosity of the solvent (water) $\eta$ is taken from Ref.~\cite{Kestin1978}.
}
\label{sup:fig:S6}
\end{figure*}

This section supplements the origin of Eq. (5) in the main text and $T_{D}$. 
Figure~\ref{sup:fig:S6} shows the temperature ($T$) dependences of $G$, $1/\eta$,  and $\Delta D$, demonstrating that they are nearly linear monotonically increasing functions of T in this narrow temperature range. 
We denote $\hat{G}(T)\equiv G(T)/G(308\,\mathrm{K})$, $1/\hat{\eta}(T)\equiv \eta(308\,\mathrm{K})/\eta(T)$, and 
$\Delta\hat{D}(T)\equiv \Delta D(T)/\Delta D(308\,\mathrm{K})$. 
In addition, we denote the approximation of these functions by the tangent line (linear function) at $T = 308$~K as follows: 
$\hat{G}(T)=\hat{a}_{G}T+(\mathrm{constant})$, 
$1/\hat{\eta}(T)=\hat{a}_{\eta}T+(\mathrm{constant})$, and
$\Delta\hat{D}(T)=\hat{a}_{\Delta D}T+(\mathrm{constant})$.
Using an approximation of Eq. (7) in the main text with neglecting the second-order terms, we have 
$\hat{a}_{\Delta D}=\hat{a}_{G}+\hat{a}_{\eta}$.
Thus, $T_D=308\,\mathrm{K}-1/\hat{a}_{\Delta D}$ is determined by $\hat{a}_{G}$ and $\hat{a}_{\eta}$.
In this case, because $(0<)\,\hat{a}_{G}\ll\hat{a}_{\eta}$ in Fig.~\ref{sup:fig:S6}, $T_D$ is mainly determined by the temperature dependence of $\hat{a}_{\eta}$ and is slightly influenced by $T_{0}$.
We remark that $1/\eta$ is a smooth monotonically increasing function of $T$ for most solvents because the viscosity behaves as $\eta \propto \exp [C_1/(T-C_2)]$ \cite{Bird2006} at temperatures farther away from the glass transition temperature.

